\documentclass[aps,prb,twocolumn,superscriptaddress]{revtex4}
\usepackage{epsfig, pslatex,latexsym,times,amssymb,amsmath,graphicx}
\usepackage{revsymb}
\usepackage{bm}
\usepackage{color}
\usepackage{amsmath}
\usepackage{amsfonts}
\usepackage{amssymb}
\usepackage{graphicx}
\providecommand{\U}[1]{\protect\rule{.1in}{.1in}}
\begin{document}
\author{N.J. Harmon}
\affiliation{Department of Physics, Ohio State University, 191 W. Woodruff Ave., Columbus, OH
43210, USA}
\author{W.O. Putikka}
\affiliation{Department of Physics, Ohio State University, 191 W. Woodruff Ave., Columbus, OH
43210, USA}
\author{R. Joynt}
\affiliation{Department of Physics, University of Wisconsin-Madison, 1150 University Ave. Madison, WI
53705, USA}
\date{\today}
\title{Theory of Electron Spin Relaxation in n-Doped Quantum Wells}

\begin{abstract}
Recent experiments have demonstrated long spin lifetimes in uniformly n-doped quantum wells. The spin dynamics of exciton, localized, and conduction spins are important for understanding these systems. We explain experimental behavior by invoking spin exchange between all spin species. 
By doing so we explain quantitatively and qualitatively the striking and unusual temperature dependence in (110)-GaAs quantum wells. 
We discuss possible future experiments to resolve the pertinent localized spin relaxation mechanisms. 
In addition, our analysis allows us to propose possible experimental scenarios that will optimize spin relaxation times in GaAs and CdTe quantum wells.

\end{abstract}
\maketitle
\section{Introduction}
In recent years, uniformly doped quantum wells (QWs) have generated increasing interest due to the long relaxation times measured therein.\cite{tribollet2, eble,chamarro}
The long relaxation times are due to spins localized on donor centers. 
While similar relaxation times have been measured in modulation doped systems, their duration has not been as reliable
due to the weaker binding energy of localized states and potential fluctuations from remote impurities.\cite{tribollet1,astakhov, zhukov}
Localization is either not seen at all\cite{tribollet1} or localization centers
thermally ionize rapidly with increasing temperature due to a small binding energy.\cite{zhukov, astakhov}
QWs uniformly doped within the well have the advantage of being characterized by well defined impurity centers with a larger 
binding energy. 
The experimental control in the amount of doping and well size make doped QWs particularly appealing to the study of quasi-two-dimensional spin dynamics.

Much of the theoretical study of spin relaxation in semiconducting systems (QWs in particular) has either focused solely on itinerant 
electrons\cite{weng, zhou,kainz} or solely on localized electrons\cite{kavokin1, kavokin2} without regard for either the presence of the other state or the 
interaction between the two states. Recently the existence and interaction between itinerant and localized states has been dealt with in bulk systems 
by Putikka and Joynt\cite{putikka} and Harmon et al.\cite{harmon}. The results of these calculations are in very good quantitative and qualitative agreement with experimental 
observations\cite{kikkawa, ghosh} in bulk n-GaAs and n-ZnO. In this paper, the theory of two interacting spin subsystems is applied to QWs.

The paper is structured as follows: 
Section \ref{section2} describes the optical generation of spin polarization in QWs; 
Section \ref{blochSection} introduces a set of modified Bloch equations to model spin dynamics;
Section \ref{occupations} calculates the equilibrium populations of localized and conductions states;
Section \ref{relaxationSection} determines the relaxation rates for all pertinent mechanisms for localized and conduction electrons;
Sections \ref{resultsGaAs} and \ref{resultsCdTe} compare our results to two GaAs QWs (uniformly doped and undoped) and one uniformly doped CdTe QW;
Section \ref{discussion} discusses our findings, suggests future work, and proposes QWs for spin liftime optimization;
we conclude in Section \ref{conclusion}. 

\section{Spin Polarization in Quantum Wells}\label{section2}
In QWs at low temperatures the creation of non-zero spin polarization, in the conduction band and donor states, proceeds from the formation of
trions (charged excitons, $X^{\pm}$) and exciton-bound-donor complexes ($D^0 X$) respectively, from the absorption of circularly polarized light. 

Polarization via the trion avenue is most relevant for modulation doped QWs where donor centers in the well are sparse.\cite{tribollet1, chamarro}
Due to the modulation doping outside the well, the number of conduction electrons in the well may be plentiful.
In such cases, assuming incident $\sigma^+$ pump pulse, a $+\frac{3}{2}$ hole and $-\frac{1}{2}$ electron are created. These bind with a resident electron from
the electron gas in the QW to form a trion ($X^-_{3/2}$). The `stolen' electron will be $+\frac{1}{2}$ to form a singlet state with the exciton's electron.
Hence, the electron gas will be left negatively polarized since the excitons are preferentially formed with spin up resident electrons. 
If the hole spin relaxes faster than the trion decays the electron gas will remain polarized.\cite{tribollet1} Selection rules dictate $+\frac{3}{2}$ ($-\frac{3}{2}$) holes 
will recombine only with $-\frac{1}{2}$
($+\frac{1}{2}$) electrons. Therefore if the hole spins relax rapidly, the released electrons will have no net polarization and the polarized electron gas will remain
predominantly negatively oriented.

A very similar picture is given for the polarization of donor bound electrons in uniformly doped QWs where the donor bound electrons play the role of the
 resident electrons.
\cite{tribollet2, eble} At low temperatures the donors are nearly all occupied and the density of the electron gas will be negligible.
When excitations are tuned at the exciton-bound-donor resonance, instead of photo-excitons binding with the resident electron gas, 
they bind with neutral donors to form the complexes $D^0 X_{3/2}$. This notation implies that
a $+\frac{3}{2}$ hole -
$-\frac{1}{2}$ electron exciton bound to a $+\frac{1}{2}$ donor bound electron. Once again for very short hole relaxation times, the donor bound electrons can be spin
polarized. 

The measured long spin relaxation times in uniformly doped QWs imply that spin polarization remains after short time processes such as
$X$ and $D^0 X$ recombination have completed. In other words, the translational degrees of freedom thermalize much more quickly than the spin degrees of freedom. 
The occupational statistics of itinerant and localized electrons are important and can be determined from equilibrium thermodynamics. 

As the temperature is increased, the electrons bound to donors thermally ionize and become itinerant. 
In analogy with the trion case, 
if the excitation energy is maintained at the $D^0 X$ frequency, the initial polarization should decrease as there are fewer $D^0 X$ complexes allowed.\cite{zhukov}
However as the number of electrons in the conduction states increases, the spin that exists on the donors will equilibrate by cross relaxing to 
conduction states by the isotropic exchange
interaction. 
If cross relaxation is rapid enough, the total spin, which is conserved by exchange, will now exist 
in the donor and conduction states weighted by their respective equilibrium densities.\cite{putikka, mahan, harmon} The polarized electron moments will then
 proceed to relax via different processes
for the localized and itinerant states. 
Since trion binding energies ($\sim 2$ meV)\cite{zhukov, astakhov} are smaller than donor-exciton binding energies ($\sim 4.5$ meV),\cite{kheng}
polarization of itinerant electrons via trion formation should be negligible as the temperature is increased. 

The above description is complicated when the photoexcitation energy is at the exciton resonance and not the exciton-bound-donor resonance. 
In such a case, the excitons may
recombine or the electron-in-exciton spin may relax before binding to a donor so one expects the low temperature spin relaxation to 
reflect also the exciton spin dynamics instead of the donor electron spin dynamics alone.\cite{zhukov} 
In essence, the electrons in an exciton represent a third spin environment with a characteristic spin relaxation time scale different 
from that of the localized donor and itinerant electrons. Because of the electron's proximity to a hole, relaxation may result from spin exchange or recombination. 

Therefore to understand the spin dynamics in QWs, it is imperative to examine the relaxation processes that affect the polarized spin moments
of the various spin systems.

\section{Modified Bloch Equations}\label{blochSection}

After rapid exciton-donor-bound complex formation, recombination, and hole relaxation, we model the zero field spin dynamics of the system in terms of modified Bloch equations:
\begin{align}
\label{bloch0}\frac{d m_{c}}{dt}  &  = -\Big( \frac{1}{\tau_{c}} + \frac
{n_{l}}{\gamma^{cr}_{c,l}} \Big) m_{c} + \frac{n_{c}}{\gamma^{cr}_{c,l}} m_{l}\nonumber\\
\frac{d m_{l}}{dt}  &  = \frac{n_{l}}{\gamma^{cr}_{c,l}} m_{c} -\Big( \frac{1}%
{\tau_{l}} + \frac{n_{c}}{\gamma^{cr}_{c,l}} \Big) m_{l}
\end{align}
where $m_c$ ($m_l$) are the conduction (localized) magnetizations, $n_c$ ($n_l$) are the conduction (localized) equilibrium occupation densities, 
$\tau_c$ ($\tau_l$) are the conduction (localized) spin relaxation times, and $\gamma^{cr}_{c,l}$ is a parameter describing the cross relaxation time between the two 
spin subsystems. Mahan and Woodworth\cite{mahan} have shown the cross relaxation time between impurity and conduction electron spins to be much shorter than any of the other spin relaxation times relevant here. We shall assume below that the same is true for the cross relaxation between electrons bound in an exciton and conduction or impurity electron spins. The motivation of these modified Bloch equations is set forth in Refs. (\onlinecite{putikka}) and (\onlinecite{harmon}). 
Eqs. (\ref{bloch0}) is valid for photoexcitation energies that do not cause free exciton formation (only two relevant spin systems). It is important to note that Eqs. (\ref{bloch0}) hold only for intermediate time scales. These scales are long compared with laser pulse times, energy relaxation times that determine subsystem populations and donor-bound exciton formation times. Fortunately, these intermediate time scales are the ones probed in the experiments.

Standard methods can be used to solve these differential equations with initial conditions $m_c(0)$ and $m_l(0)$. We assume that the initial spin polarization is perpendicular to the QW's growth plane and that the excitation density, $N_x$, is small enough such that the resultant spin relaxation time, $\tau_s$, will not depend strongly on $N_x$.\cite{eble} The solutions yield a time dependence of the total magnetization
$m(t) = m_c(t) + m_l(t)$ to be a sum of two exponentials - one of which is $\exp(-t/\tau_s)$  and the other of which has a time constant proportional to the cross
relaxation time.
In the case of rapid cross relaxation (faster than all spin relaxation mechanisms), only one exponential survives and we express the total relaxation rate as
\begin{equation}\label{bloch}
\frac{1}{\tau_s} = \frac{n_l}{n_{imp}} \frac{1}{\tau_l} + \frac{n_c}{n_{imp}} \frac{1}{\tau_c}
\end{equation}
where $n_{imp} = n_l + n_c$ is the total impurity concentration. 
This model, or variations of it, has been successfully applied to bulk n-GaAs and bulk n-ZnO. \cite{putikka, harmon}

If the photoexcitation energy is set near the exciton energy, the Bloch equations must be modified to take into account exciton spin relaxation and multiple cross relaxations: $\gamma_{i,j}$ for $i,j \in c, l, x$ for conduction, localized, and excition spins respectively.
We model exciton spin relaxation as electron-in-exciton spin relaxation\cite{adachi2} and assume that hole spin relaxation is very rapid. Eq. (\ref{bloch0}) generalizes to
\begin{align}
\label{bloch2}
\frac{d m_{c}}{dt}  &  = -\Big( \frac{1}{\tau_{c}} + \frac
{n_{l}}{\gamma^{cr}_{c,l}}+\frac{n_x}{\gamma^{cr}_{c,x}} \Big) m_{c} + \frac{n_{c}+N_x - n_x}{\gamma^{cr}_{c,l}} m_{l} + \frac{n_{c}+N_x - n_x}{\gamma^{cr}_{c,x}} m_{x} \nonumber\\
\frac{d m_{l}}{dt}  &  = \frac{n_{l}}{\gamma^{cr}_{c,l}} m_{c} -\Big( \frac{1}%
{\tau_{l}} + \frac{n_{c} + N_x - n_x}{\gamma^{cr}_{c,l}}+\frac{n_x}{\gamma^{cr}_{l,x}} \Big) m_{l} + \frac{n_{l}}{\gamma^{cr}_{l,x}} m_{x} \nonumber\\
\frac{d m_{x}}{dt}  &  = \frac{n_x}{\gamma^{cr}_{c,x}} m_{c} + \frac{n_x}{\gamma^{cr}_{l,x}} m_{l} -\Big( \frac{1}%
{\tau_{x}} + \frac{n_{c} + N_x - n_x}{\gamma^{cr}_{c,x}} + \frac{n_l}{\gamma^{cr}_{l,x}} \Big) m_x,
\end{align}
where $\tau_x$ represents spin lifetime of an electron bound to a hole. $n_x$ ($m_x$) is the number (magnetization) of electrons bound in an exciton. 
$N_x$ is the initial density of photoexcited electrons and the quantity $N_x - n_x$ is the number of photoexcited electrons that do not participate in an exciton.
We assume quasi-equilibrium such that $n_x$ is determined from thermodynamics (see Section \ref{occupations}). It should be stated that Eq. (\ref{bloch2}) is valid only for
times shorter than the recombination time; in other words, on a time scale where $N_x$ can be assumed to not change significantly. 
Recombination times have been measured\cite{adachi} in similar systems as to those studied here to be longer than the observed spin relaxation times so this
approximation seems justified. 
In Section \ref{resultsGaAs}, we find that the effects of recombination of free carriers can be added to $1/\tau_c$ to obtain excellent agreement with the experimental data.

If we solve the system of equations in Eq. (\ref{bloch2}) as we did for Eq. (\ref{bloch0}), we obtain the relaxation rate
\begin{equation}\label{bloch3}
\frac{1}{\tau_s} = \frac{n_l}{n_{imp}+N_x} \frac{1}{\tau_l} + \frac{n_c+N_x-n_x}{n_{imp}+N_x} \frac{1}{\tau_c}  + \frac{n_x}{n_{imp}+N_x} \frac{1}{\tau_x}.
\end{equation}

For both Eqs. (\ref{bloch}) and (\ref{bloch3}), we allow $\tau_l$, $\tau_c$, and $\tau_x$ to be phenomenological parameters of the form 
$\tau_i^{-1} = \sum_j 1/\tau_j$ where $j$ refers to a type of spin relaxation mechanism. 
From the experimental constraints and results, we can determine which relaxation mechanisms are important.

\section{Occupation Concentrations}\label{occupations}
As shown above, the relative occupations of localized and itinerant states play an important role in our theory.
Fortunately, in two dimensional systems, the occupation probabilities of the two states ($n_l/n_{imp}$ and $n_c/n_{imp}$) can be determined
exactly. The densities we are interested in are dilute enough such that the non-degenerate
limit (Boltzmann statistics) can be utilized.

	\begin{figure}[hbtp]\label{occ}
	\begin{centering}
	\includegraphics[scale = 0.32,trim = 0 00 00 00, angle = 0,clip]{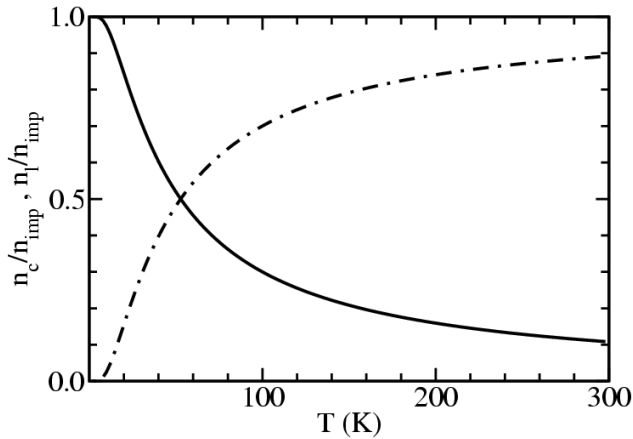}
	\caption[]{Occupation probabilities of localized (solid line) and conduction (dash-dotted line) states with impurity density $n_{imp} = 4 \times 10^{10}$ cm$^{-2}$ determined from Eqs. (\ref{oc1}, \ref{oc2}). Other parameters
	for GaAs are $a_B^{\ast} = 10.4$ nm and $m^{\ast} = 0.067 m$.}
	\end{centering}
	\end{figure}

The probability for a donor to be singly occupied (only the ground state needs to be considered\cite{look}) is\cite{ashcroft}
\begin{equation}
\frac{n_l}{n_{imp}} = \frac{1}{\frac{1}{2} e^{(E_B - \mu)/k_B T} + 1}.
\end{equation}
The density of itinerant states is given by
\begin{equation}
n_c = N_c  e^{ \mu/k_B T}
\end{equation}
where $N_c = m^{\ast} k_B T / \hbar^2 \pi$ and the conduction band edge is taken to be zero energy.
The chemical potential $\mu$ can be found using the constraint
\begin{equation}\label{oc1}
\frac{n_l}{n_{imp}}+\frac{n_c}{n_{imp}}   = 1.
\end{equation}
Using the result for $\mu$, one obtains
\begin{equation}\label{oc2}
\frac{n_l}{n_{imp}} = \frac{\sqrt{1+Q(T, n_{imp})}-1}{\sqrt{1+Q(T, n_{imp})}+1},
\end{equation}
where
\begin{equation}
Q(T,n_{imp}) = \frac{8 n_{imp}}{N_c} e^{-E_b/k_B T}.
\end{equation}

An example of the temperature dependence of these occupation probabilities is shown for a GaAs QW in Figure 1 where $n_{imp} = 4 \times 10^{10}$ cm$^{-2}$.
At the lowest temperatures, the donors are fully occupied. As the temperature increases, $n_l$ decreases and $n_c$ increases to where at around $50$ K, the two occupation probabilities
are equal. From Eqs. (\ref{bloch}, \ref{bloch3}), it is evident that these occupational statistics have ramifications in the measured spin relaxation times.
The results here are also applied to the excitons in quasi-equilibrium.

\section{Spin Relaxation}\label{relaxationSection}

We now discuss the relevant spin relaxation mechanisms for both localized and conduction electrons. 
The electron-in-exciton spin relaxation, $\tau_x$, is a combination of electron-hole recombination and electron-hole exchange relaxation.
Due to its complicated nature we defer the calculation of $\tau_x$ to future work. Here we treat it as a phenomenological parameter.

\subsection{Localized Spin Relaxation}	

First we discuss spin relaxation via the anisotropic spin exchange for donor bound electrons. This has been treated extensively elsewhere.
\cite{dzyaloshinskii, moriya, gorkov2, kavokin1} Most recently it has been examined by Kavokin in Ref. (\onlinecite{kavokin2}). It is his treatment that we 
detail below for semiconducting QWs.

Kavokin argues\cite{kavokin2} that some portion of localized relaxation results from spin diffusion due to the exchange interaction between donors. Anisotropic 
corrections to the isotropic exchange Hamiltonian cause a spin to rotate through an angle $\gamma_{i,j}$ when it is transferred between two donor centers located at 
positions $r_i$ and $r_j$. 
The angle-averaged rotation angle is $\langle \gamma_{i,j}^2 \rangle ^{1/2} = \langle r_{i,j}^2 \rangle ^{1/2}/L_{s.o.}$ where $L_{s.o.}$ is the spin orbit length.\cite{kavokin2}
The spin is relaxed when the accumulated rotation angle $\Gamma$ becomes on the order of unity such that $\Gamma^2 =\sum \langle \gamma_{i,j} ^2
\rangle = \sum \langle r_{i,j}^2 \rangle/L_{s.o.}^2 = 2 D_{ex} \tau_{ex}/L_{s.o.}^2 = 1$ where $D_{ex}$ is the diffusion coefficient and
the relaxation time is 
\begin{equation}
\tau_{ex} =\frac{L^2_{so}}{2 D_{ex}}.
\end{equation}
In quasi-2D (100) QWs where Dresselhaus bulk inversion asymmetry (BIA) terms dominate,\cite{kavokin1}
\begin{equation}
L_{s.o.}= \Big(\frac{2 \alpha \hbar}{\sqrt{2 m^* E_g}} \langle k_z^2 \rangle\Big)^{-1},
\end{equation}
where $\alpha$ is a dimensionless measure of the spin orbit strength and $\langle k_z^2 \rangle$ is due to the quasi-2D confinement and is of the form $\beta^2/L^2$. For infinite well confinement $\beta = \pi$.
The diffusion coefficient is approximately\cite{kavokin2}
\begin{equation}
D_{ex}= \frac{1}{2}\langle r_{i,j}^2 \rangle \langle J \rangle/\hbar.
\end{equation}
with exchange constant\cite{ponomarev} in 2D
\begin{equation}
J_{2D}= 15.21 E_b \Big(\frac{r_{i,j}}{a_B}\Big)^{7/4} e^{-4 r_{i,j}/a_B}
\end{equation}
where $E_b$ is the binding energy: $E_b = \hbar^2/(2 m^* a_B^2)$.
How $r_{i,j}$ is to be determined will be discussed in Section \ref{resultsGaAs}.

These results can be combined to obtain the relaxation rate in terms of a dimensionless impurity separation scale, $x$:
\begin{equation}\label{dm2}
\frac{1}{\tau_{ex}} =15.21 \frac{\alpha^2 \hbar^3 \langle k_z^2 \rangle^2}{E_g {m^{\ast}}^2}\langle x^2 \rangle \langle x^{7/4} e^{-4 x} \rangle
\end{equation}
where $x= r_{i,j}/a_B$.

Localized electron spins may also relax due to nuclear fields. A localized electron is coupled to many nuclear spins by the hyperfine interaction. 
To the electrons,  these nuclear spins appear as a randomly fluctuating field but these nuclear fields can be assumed quasi-stationary since the nuclear evolution
time is much longer than electron evolution time due to the contrast in magnetic moments.\cite{kavokin2} 
What governs the electron spin evolution is the electron correlation time, $\tau_{corr}$. 
If $\tau_{corr}$ is long such that $\mu_B g^{\ast}/\hbar \langle B_N^2 \rangle ^{1/2} \tau_{corr}> 1$ (where $\langle B_N^2 \rangle ^{1/2} = 
B_N^{max}/\sqrt{N_{L}}$ is the root-mean-square field, $B_N^{max}$ is the maximum nuclear field, and $N_{L}$ is the number of nuclei in the electron's localization
volume), 
then the electron polarization decays due to ensemble
dephasing; there will be random electron precession frequencies due to a random distribution of frozen nuclear fields.\cite{merkulov}
If the mechanism contributing to the electron correlation time is exchange induced spin diffusion, 
$\tau_{corr}$ is estimated to be $(n_{imp}^{1/2} D_{ex})^{-1}$ in quasi-two dimensions.\cite{kavokin2}

Merkulov et al.\cite{merkulov} find a dephasing rate for quantum dots to be
\begin{equation}\label{hyperfine}
\frac{1}{\tau_{nuc}} = \sqrt{\frac{16 \sum_j I_j (I_j+1) A_j^2}{3 \hbar^2 N_L}}
\end{equation}
where the sum over $j$ is a sum over all nuclei in the unit cell, $I_j$ is the nuclear spin, $A_j$ is the hyperfine constant, and $N_L$ is the number
of nuclei in the electron's localized volume. 
It is important to state that this spin dephasing does not decay exponentially but decreases to $10\%$ of the original spin polarization in $\tau_{nuc}$ 
and then increases to $33\%$ of the original spin polarization in $2 \tau_{nuc}$ where 
it will then decay at a much slower rate.\cite{merkulov, braun}

If $\tau_{corr}$ is short such that $\mu_B g^{\ast}/\hbar \langle B_N^2 \rangle ^{1/2} \tau_{corr} \ll 1$, then the relaxation will be of the motional narrowing type.\cite{kavokin2}

\subsection{Conduction Spin Relaxation}	
Conduction band states undergo ordinary impurity and phonon scattering. Each scattering event gives a change in the wave vector $\mathbf{k}$, which in turn
changes the effective magnetic field on the spin that comes from spin-orbit coupling. This fluctuating field relaxes the spin. This is known as the D'yakonov-Perel' (DP) spin relaxation mechanism.\cite{dyakonov1, dyakonov2} The effective field strength 
is proportional to the conduction band splitting. 
In this article, we are interested in conduction spin relaxation in (001) and (110) oriented QWs.
For (001) QWs the spin relaxation rate results from a spin-orbit term in the Hamiltonian,
 $H_{s.o} =\frac{\hbar}{2} \Omega(\mathbf{k_{||}})\cdot \mathbf{\sigma}$ where\cite{kainz}
\begin{displaymath}
\Omega(\mathbf{k_{||}})= \frac{2 \gamma}{\hbar} 
\left(\begin{array}{c}
k_x (k_y^2 - \langle k_z^2 \rangle)\\
k_y (\langle k_z^2 \rangle - k_x^2)\\
0
\end{array}\right).
\end{displaymath}
The angular brackets denote spatial averaging across the well width. $\gamma$ is a band parameter that governs the magnitude of the spin-orbit splitting. For GaAs, $\gamma \sim 17$ meV nm$^3$.\cite{fu} We assume the QWs have been grown symmetrically and therefore ignore any 
Rashba contribution.\cite{rashba}

The resulting spin relaxation has been worked out in detail by Kainz et al. in Ref. (\onlinecite{kainz}). For the experiment\cite{ohno} we compare to, we find the non-degenerate limit to be 
applicable and hence use the relaxation rate for spin oriented in the z-direction, 
\begin{eqnarray}\label{DP}
\frac{1}{\tau_z} &=& 
\frac{4}{\hbar^2} \tau_p(T) \Bigg[\gamma^2 \langle k_z^2 \rangle^2 \frac{2 m^{\ast}k_B T}{\hbar^2} k_B T-
\frac{\gamma^2 \langle k_z^2 \rangle}{2} \Big(\frac{2 m^{\ast}k_B T}{\hbar^2 }\Big)^2 j_2 +{}\nonumber\\
& & {}\gamma^2 \frac{1+\tau_3/\tau_1}{16} \Big(\frac{2 m^{\ast}k_B T}{\hbar^2 }\Big)^3 j_3\Bigg]
\end{eqnarray}
where $j_2 \approx 2$ and $j_3 \approx 6$ depend on the type of scattering mechanism. 
We assume Type I scattering as defined in Ref. (\onlinecite{kainz}).
The ratio $\tau_3/\tau_1$ is unity for Type I scattering. $\tau_p(T)$ is the momentum relaxation time which can be extracted
from mobility measurements.

A more interesting case is that of (110) QWs where the spin-orbit Hamiltonian is\cite{hassenkam}
\begin{equation}
H_{s.o.} = -\gamma \mathbf{\sigma}_z k_x \big(\frac{1}{2}\langle k_z^2 \rangle - \frac{1}{2} (k_x^2 - 2 k_y^2) \big)
\end{equation}
which is obtained from the (001) Hamiltonian by transforming  the coordinate system such that $x||[\overline{1}10]$, $y||[001]$, and $z||[110]$. As can be seen from
 the form of this Hamiltonian, the effective magnetic field is in the direction of the growth plane. Hence, spins oriented along the effective field will experience 
 no spin relaxation.

Conduction spins also relax due to the Elliott-Yafet (EY) mechanism\cite{elliott, yafet} which arises from spin mixing in the wavefunctions. Due to spin-orbit
interaction, when a conduction electron is scattered by a spin-independent potential from state $\mathbf{k}$ to $\mathbf{k}'$, the initial and final
states are not eigenstates of the spin projection operator $S_z$ so the process relaxes the spin. In bulk, the relaxation rate is known to be of the form 
$1/\tau_{EY} = \alpha_{EY} T^2/\tau_p(T)$ where $\alpha_{EY}$ is a material-dependent parameter and $\tau_p$ is the momentum relaxation time.\cite{chazalviel}

However the EY mechansim in quasi-two dimensions will not take the same form since $\mathbf{k}$ will be quantized in one direction (the direction of confinement). 
The treatment in bulk\cite{ridley} has been extended to QWs to obtain\cite{tackeuchi}
\begin{equation}
\frac{1}{\tau_{EY}} \approx  \Big(\frac{\Delta_{s.o.}}{\Delta_{s.o.}+E_g}\Big)^2 \Big(1-\frac{m^{\ast}}{m}\Big)^2 \frac{E_c k_B T}{E_g^2} \frac{1}{\tau_p(T)},
\end{equation}
where $\Delta_{s.o.}$ is the spin-orbit splitting energy and $E_c$ is the QW confinement energy.

Spins may also relax due to the Bir-Aronov-Pikus (BAP) mechanism\cite{bir} which arises from the scattering of electrons and holes. 
This relaxation mechansim 
is commonly considered efficient only in $p$-type materials when the number of holes is large.\cite{song}
We fit the experimental data in Section \ref{resultsGaAs} without consideration of this mechanism.

We will now examine how these relaxation mechanisms are manifest in two different QWs. 

\section{Results for G\MakeLowercase{a}A\MakeLowercase{s}/A\MakeLowercase{l}G\MakeLowercase{a}A\MakeLowercase{s} Quantum Well}\label{resultsGaAs}

We apply our method to measured spin relaxation times of two GaAs/AlGaAs QWs by Ohno et al.: \cite{ohno, ohno2}
(100) n-doped QW with doping $n_{imp} = 4 \times 10^{10}$ cm$^{-2}$, well width $L = 7.5$ nm;
and a (110) undoped QW with well width $L = 7.5$ nm.
In both (pump-probe) experiments, the pump or photoexcitation energy was tuned to the heavy hole exciton resonance and normally incident on the sample.
As mentioned in Section \ref{section2}, the exciton spin becomes important at low temperatures for such excitation energies.
The experimental spin relaxation times as a function of temperature are displayed (solid circles) in Figures 2 and 3.\cite{footnote}

	\begin{figure}[hbtp]\label{110}
	\begin{centering}
	\includegraphics[scale = 0.32,trim = 0 00 00 00, angle = 0,clip]{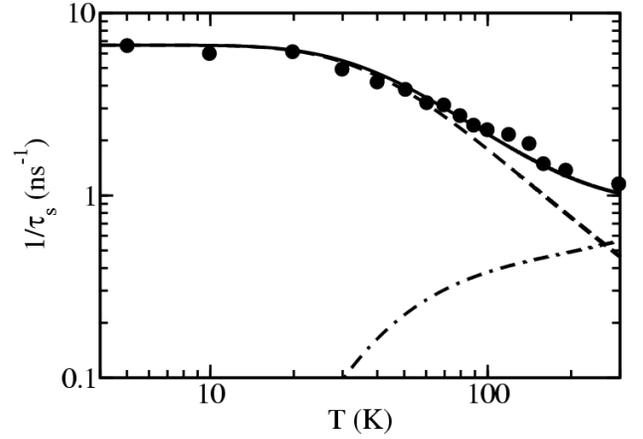}
	\caption[]{Spin relaxation versus temperature in undoped (110) GaAs QW. Points are experiment of Ref. (\onlinecite{ohno2}). 
	Dash-dotted line: Using only conduction portion of 
	Eq. (19)  and $1/\tau_c = 1/\tau_{EY} + 1/\tau_r$. Dashed line:  using only excitonic portion of 
	Eq. (19).
	Solid line: Eq. (19). Spin relaxation rate of excitons decreases with temperature increase due to thermal ionization. Conduction spin relaxation is longer in (110) QW than in other oriented QWs due to vanishing DP mechanism.}
	\end{centering}
	\end{figure}

For the undoped (110) QW, Eq. (4) is modified to become
\begin{equation}\label{bloch4}
\frac{1}{\tau_s} = \frac{N_x-n_x}{N_x} \frac{1}{\tau_c}  + \frac{n_x}{N_x} \frac{1}{\tau_x}.
\end{equation}
For this sample, at low temperatures, $n_x = N_x$ so the $\tau_s = \tau_x \approx 0.15$ ns. 
At higher temperatures, recombination (in time $\tau_r$) and EY act to relax conduction spins since DP relaxation is significantly reduced for the (110) QW orientation. 
To account for the quasi-two dimensional nature of the QW, we use an intermediate value (between 2D and 3D values) for the exciton's binding energy.\cite{harrison}
Eq. (\ref{bloch4}) (solid line)  fits the data (points) with excellent agreement in Figure 2 when
$N_x = 1.5 \times 10^{10}$ cm$^{-2}$ and $\tau_r = 2$ ns which are near the experimentally reported values\cite{adachi} ($N_x \approx 10^{10}$ cm$^{-2}$ and $\tau_r \approx 1.6$ ns).
The contributions from the excitons and conduction electrons are also shown (dashed and dash-dotted lines respectively).
The trend in the data is well described by our theory - at low temperatures excitons predominate and the spin relaxation time is $\tau_x$. 
When the temperature increases, the excitons thermally ionize leading to net moment in the conduction band. 
Since the conduction  band spin relaxation time is longer than the exciton spin relaxation time, the measured relaxation time increases with temperature as
described in Eq. (\ref{bloch4}). We expect the relaxation times to eventually level out as the excitons disappear. 
Eventually, the relaxation time will decrease as the temperature dependence of EY takes effect.

For the doped (100) QW, Eq. (4) should be used to describe the temperature dependence of the relaxation rate. 
Using the values from above and $n_{imp} = 4 \times 10^{10}$ cm$^{-2}$, $\tau_s = 0.35$ ns, we can extract the approximate value of $\tau_l$. 
In doing so we obtain $\tau_l \approx 0.5$ ns. We stress that this value has considerable uncertainty due to the uncertainty in the parameters (namely $N_{x}$) that determine $\tau_l$. 
The presence of impurities has lengthened the observed low temperature spin relaxation time by more than a factor of two. The relaxation 
time in the doped sample can be further increased by reducing the excitation density. 
As the temperature is increased, donors become unoccupied and conduction electrons will play a larger role in relaxation as expressed in Eq. (4).
We can determine the main conduction spin relaxation mechanism by investigating its temperature dependence.

	\begin{figure}[hbtp]\label{gaasQW}
	\begin{centering}
	\includegraphics[scale = 0.32,trim = 0 00 00 00, angle = 0,clip]{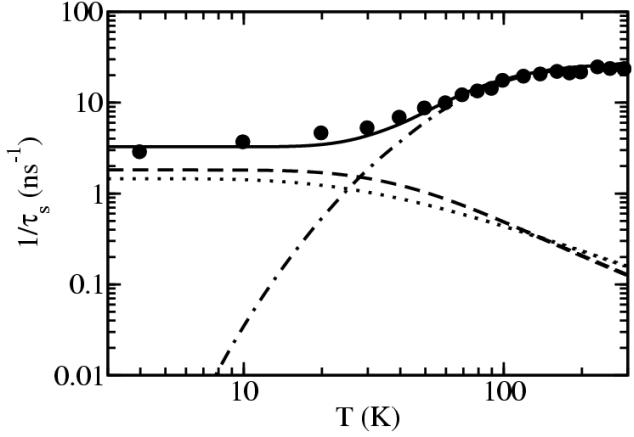}
	\caption[]{Spin relaxation versus temperature in n-doped (100) GaAs QW.
	Points are from Ref. (\onlinecite{ohno}). Dashed line: excitonic contribution in Eq. (4). Dotted line: localized contribution in Eq. (4).
	Dash-dotted line: conduction contribution in Eq. (4). Solid black line: Eq. (4). Both exciton and localized spin relaxation contribute to the observed low temperature spin relaxation. Conduction spin relaxation is the most strong contributor to the observed relaxation at higher temperatures. }
	\end{centering}
	\end{figure}

We are now left with the task of determining what the localized and conduction spin relaxation mechanisms are.
We plot the relaxation rate for the n-doped GaAs QW as a function of temperature in Figure 3. 
The dashed, dotted, and dash-dotted lines refer to the three terms of Eq. (4) - the density weighted average of the respective relaxation rates.
The solid line is the sum of all three terms. 

We begin by calculating spin relaxation due to spin exchange diffusion in Eq. (\ref{dm2}).
This is difficult due to the exponential dependence on $r_{i,j}$. 
For GaAs, $\alpha = 0.06$, $E_g = 1.52$ eV, $m^{\ast} = 0.067 m$, and $a_B = 10.4$ nm.
To calculate $ \langle k_z^2 \rangle = \beta^2/L^2$, we need to know the band offsets and assume a finite square well. 
The potential depth for a AlGaAs QW is about $V_0 = 0.23$
eV. This comes from $\frac{\Delta E_c}{\Delta E_g} = 0.62$ and $\Delta E_g = 0.37$ eV in GaAs.\cite{davies} 
From this we can determine $\beta$ which will also depend on the well width $L$. 
For $L = 7.5$ nm, $\beta = 2.19$. Of course in the limit of $V_0 \rightarrow \infty$, $\beta \rightarrow \pi$. 
What remains to be determined is $r_{i,j}$ which is proportional to the inter-donor separation $r_{i,j} = \gamma n_{imp}^{-1/2}$.
For average inter-donor spacing in two dimensions, $\gamma_{av} = 0.564$.
When we allow $\gamma$ to be fitting parameter, we obtain $r_{i,j} = 19.5$ nm which corresponds to $\gamma = 0.4$. 

We now determine the relaxation rate due to the hyperfine interaction. 
Since $\mu_B g^{\ast}/\hbar \langle B_N^2 \rangle ^{1/2} \tau_{corr} \gg 1$ when $n_{imp} = 4 \times 10^{10}$ cm$^{-2}$, the hyperfine relaxation is described by Eq. (\ref{hyperfine}). 
Since nearly all nuclei have the same spin\cite{schliemann} ($I = 3/2$), we can express Eq. (\ref{hyperfine}) as
\begin{equation}
\frac{1}{\tau_{nuc}} = 2 \sqrt{\frac{5\sum_j  A_j^2}{\hbar^2 N_L}},
\end{equation}
with 
$\sum_j  A_j^2 = 1.2 \times 10^{-3}$ meV$^2$ and $N_L \sim 2.1 \times 10^5$.\cite{merkulov} This yields $\tau_{nuc} = 3.9$ ns.
Due to the donor's confinement in the QW, its wavefunction may shrink thereby reducing the localization volume and therefore also reducing $N_L$ and $\tau_{nuc}$ .\cite{harrison}

In Figure 3, we find find excellent agreement with experiment over a large temperature range when $\tau_p(T)$ in Eq. (\ref{DP}) is made a factor of three smaller than what is reported in Ref. (\onlinecite{kainz}).
We attain approximately the same quantitative accuracy as in Ref. (\onlinecite{kainz}) but since we also take into account the localized spins, we find excellent qualitative agreement
as well.
It should be emphasized that the quadratic and cubic terms of Eq. (\ref{DP}) are important in the high temperature regime. 
The EY rate is qualitatively and quantitatively different from the data. For instance, $1/\tau_{EY} \approx 0.1$ ns$^{-1}$ at 300 K so we rule it out of contention. 
We also now ignore recombination of carriers since an appreciable amount of 
equilibrium carriers exist (n-doped system) leading to recombination of primarily non-polarized spins.

One would not expect these results to agree with spin relaxation measurements in modulation doped QWs.
In modulation doped systems, the occupation densities $n_l$ and $n_c$ cannot be calculated as we have done here.
In such systems different spin relaxation dependencies are seen.\cite{adachi, dohrmann}

\section{Results for C\MakeLowercase{d}T\MakeLowercase{e}/C\MakeLowercase{d}M\MakeLowercase{g}T\MakeLowercase{e} Quantum Well}\label{resultsCdTe}
 
The experiment by Tribollet et. al. on a n-CdTe QW offers an instructive complement to the previous experiments on GaAs.
In their experiment, Tribollet et al. measure spin relaxation times $\tau_s \approx 20$ ns for CdTe/CdMgTe QWs with $n_{imp} = 1 \times 10^{11}$ cm$^{-2}$.
Importantly, they excited with laser energies at the donor bound exciton frequency instead of the heavy hole exciton frequency. 

For CdTe, $E_g = 1.61$ eV, $m^{\ast} = 0.11 m$, and $a_B = 5.3$ nm. The spin-orbit parameter, $\alpha$ is not known but we approximate it by noting that 
the spin-orbit splitting energy in CdTe is $\Delta_{s.o} = 0.927$ eV whereas in GaAs, it is $\Delta_{s.o} = 0.34$ eV.
Since $\alpha$ is approximately proportional to $\Delta_{s.o}$, we obtain $\alpha = 0.164$ for CdTe.

To obtain potential well depth for CdTe QW, use $E_g(x_{Mg}) = 1.61+1.76 x_{Mg}$ where $x_{Mg}$ gives fraction of Mg in Cd$_{1-x}$Mg$_{x}$Te.\cite{zaitsev} 
If we use $x_{Mg}=0.1$, we get $V_0 = 0.12$eV which leads to $\beta = 2.18$.

We now determine the relaxation rate due to the hyperfine interaction. 
Since all nuclei with non-zero spin will have the same spin\cite{schliemann} ($I = 1/2$), we can express Eq. (\ref{hyperfine}) as
\begin{equation}
\frac{1}{\tau_{nuc}} = 2 \sqrt{\frac{\sum_j  A_j^2 P_j}{\hbar^2 N_L}},
\end{equation}
where $P_j$ has been addended to account for isotopic abundances.\cite{tribollet2} The natural abundancies of spin-1/2 Cd and Te nuclei dictate that $P_{Cd} = 0.25$ and $P_{Te} =0.08$. 
The remaining isotopes are spin-0. $N_L = 1.8 \times 10^{4}$, 
$A_{\textrm{Cd}} = 31$ $\mu$eV, and $A_{\textrm{Te}} = 45$ $\mu$eV which yields $\tau_{nuc} = 4.4$ ns.\cite{tribollet2}
The confined donor wavefunction in CdTe should shrink less than in GaAs since the effective Bohr radius is half as large.

We see that this value is within an order of magnitude of what we have calculated for relaxation due to the hyperfine interaction. 
We can also compare the experimental time to what we obtain for spin exchange diffusion.
When we allow $\gamma$ to be a fitting parameter, we obtain $r_{i,j} = 19.3$ nm which corresponds to $\gamma = 0.61$. This is in reasonable agreement with $\gamma_{av}$.

Unfortunately no relaxation measurements have been performed at higher temperatures in n-doped CdTe QWs that we are aware of. 
We are also not aware of mobility measurements in n-doped CdTe QWs. 
The prevalent mechansim (DP or EY) will depend on the mobility so we forgo determining the more efficient rate. 
However, in analogy to bulk systems, we expect the CdTe QW mobilities to be less than 
the GaAs QW mobilities.\cite{putikka,segall}
In the next section we analyze CdTe's spin relaxation rate for (110) grown crystal so DP can be ignored.
	
\section{Comparison of G\MakeLowercase{a}A\MakeLowercase{s} and C\MakeLowercase{d}T\MakeLowercase{e} Quantum Wells}\label{discussion}

First we discuss the low temperature spin relaxation. 
Interestingly, the localized relaxation time in CdTe is about 20 times longer than in GaAs. 
This can be explained  by the spin exchange relaxation despite the larger spin orbit parameter in the CdTe. 
This is more than offset by the smaller effective Bohr radius in CdTe ($5.3$ nm vs. $10.4 $ nm) and the exponential behavior of the anistropic exchange relaxation. 
However due to the exponential factor, any discrepancy between the two QWs can be explained by adjusting their respective $\gamma$s appropriately, though the fitted $\gamma$s do fall near $\gamma_{avg}$. 
The discrepancy in times is difficult to explain by the hyperfine interaction since the two calculated relaxation times are very near each other. 
Additionally, no plateau effect is seen that is indicative of hyperfine dephasing.\cite{tribollet2, braun}
Another possibility is that one QW is governed by relaxation from spin exchange and the other from hyperfine interactions. 
Without experimental data, answering these questions is difficult. 
It is our hope that further experiments will be done to sort out these questions.
However, we can propose ways in which these answers can be discovered.

Relaxation by anisotropic spin exchange is strongly dependent on the impurity density. 
By altering the impurity doping within the well, one should see large changes in the spin relaxation time if this mechanism is dominant.
From Eq. (\ref{dm2}) we see that this mechansim will also depend on the confinement energy. 
Hence this mechanism should also be affected by changing the well width. 
The hyperfine dephasing mechanism should be largely unaffected by impurity concentration differences as long as they are not so extensive as to 
cause the correlation time to become very short and enter a motional narrowing regime. 
Varying the well width will have an effect on the donor wavefunctions, but as long as they are not squeezed too thin the effect should 
not be dramatic.

	\begin{figure}[hbtp]\label{temp2}
	\begin{centering}
	\includegraphics[scale = 0.32,trim = 0 00 00 00, angle = 0,clip]{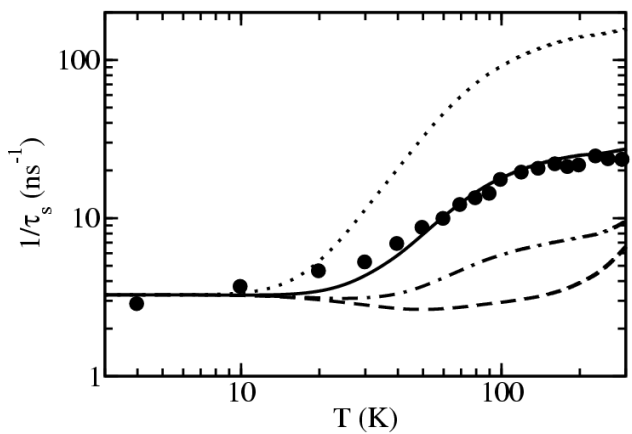}
	\caption[]{Spin relaxation in GaAs (100) QWs with different well widths (all other parameters, including $\tau_l$ and $\tau_x$, do not change). 
	Points are from Ref. (\onlinecite{ohno}) where $L_0 = 7.5$ nm. 
	Dotted: $2 L_0$; dash-dotted: $3 L_0/2$; solid: $L_0$;
	dashed: $L_0/2$.}
	\end{centering}
	\end{figure}
	
For spin relaxation at higher temperatures, DP prevails in (100) GaAs QWs as mentioned earlier. 
Whether DP or EY is more efficient in CdTe depends on the momentum relaxation time. 
By changes in momentum relaxation times (by changing well width or impurity concentration), 
we predict the the possibility to induce a clear `dip' in the temperature dependence which we see in 
Figure 4. This same non-monotonicity has been observed bulk GaAs and ZnO.\cite{kikkawa, ghosh, putikka, harmon}

Using our results we propose that n-doped (110) QWs should optimize spin lifetimes (when excited at exciton-bound-donor frequency) since DP is suppressed. 
Figure 5 displays our results for GaAs and CdTe (110) QWs as impurity densities $n_{imp} = 4 \times 10^{10}$ cm$^{-2}$ and $n_{imp} = 1 \times 10^{11}$ cm$^{-2}$
respectively. 
The decrease seen in GaAs is now due to depopulation of donor states instead of exciton thermalization.
The depopulation is much slower in CdTe since the doping is higher.
The up-turn in the CdTe curve as room temperature is reached is due to EY which is too weak to be seen in GaAs. 
We plot the data points from the undoped (110) GaAs
QW for comparison. By avoiding the creation of excitons and their short lifetimes, long spin relaxation times can be achieved.

	\begin{figure}[hbtp]\label{gaasCDTE}
	\begin{centering}
	\includegraphics[scale = 0.32,trim = 0 00 00 00, angle = 0,clip]{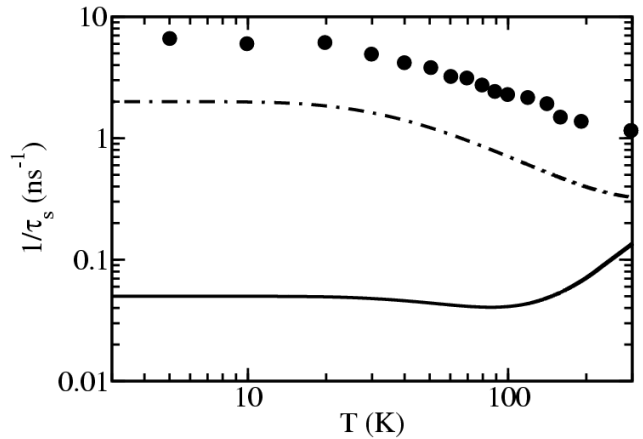}
	\caption[]{Spin relaxation in (110) GaAs ($n_{imp} = 4 \times 10^{10}$ cm$^{-2}$): dashed-dotted line. 
	Spin relaxation in (110) CdTe ($n_{imp} = 1 \times 10^{11}$ cm$^{-2}$): solid line. Points from undoped (110) GaAs QW experiment\cite{ohno2} are included for comparison. For both systems, $\tau_p(T)$ from Ref. (\onlinecite{kainz}) were used. EY is too weak over the temperature range depicted to be seen in the GaAs system.
However EY is the cause of the increase in spin relaxation rate for the CdTe system.}
	\end{centering}
	\end{figure}

\section{Conclusions}\label{conclusion}
We find that the spin relaxation times in n-doped QWs can be well described by a theory invoking spin exchange between spin species. In undoped (110) QWs, where DP is absent, we find that exciton spin relaxation is important and leads to the observed surprising temperature dependence. We predict that a similar temperature dependence (though with longer relaxation times) should be observed in n-doped (110) QWs when excited at the exciton-bound-donor frequency.

We have suggested future experimental work to resolve what mechanisms relax spin localized on donors in n-doped GaAs and CdTe QWs. The theory allows us to predict experimental conditions that should optimize the measured spin relaxation times in GaAs and CdTe QWs.

\section{Acknowledgements}

Financial support was provided by the National Science Foundation, Grant Nos.
NSF-ECS-0523918 (NH and WP) and NSF-ECS-0524253 (RJ). NH also acknowledges the Center for Emergent Materials at the Ohio State University, an NSF MRSEC (Award Number DMR-0820414), for providing partial funding for this research.\ \

\ \ \ 

\

\begin{thebibliography}{999999999999999999999999999999999999999999999999999999999999999999999999999999999999999999999999} %
\expandafter\ifx\csname natexlab\endcsname\relax

\fi
\expandafter\ifx\csname bibnamefont\endcsname\relax


\fi
\expandafter\ifx\csname bibfnamefont\endcsname\relax


\fi
\expandafter\ifx\csname citenamefont\endcsname\relax


\fi
\expandafter\ifx\csname url\endcsname\relax


\fi
\expandafter\ifx\csname urlprefix\endcsname\relax

\fi
\providecommand{\bibinfo}[2]{#2} \providecommand{\eprint}[2][]{\url{#2}}

\bibitem[Tribollet et~al.(2007)Tribollet, Aubry, Karczewski, Sermage, Bernardot,Testelin, and Chamarro]%
{tribollet2}\bibinfo{author}
\bibinfo{author}{\bibfnamefont{J.}~\bibnamefont{Tribollet}},
\bibinfo{author}{\bibfnamefont{E.}~\bibnamefont{Aubry}},
\bibinfo{author}{\bibfnamefont{G.}~\bibnamefont{Karczewski}},
\bibinfo{author}{\bibfnamefont{B.}~\bibnamefont{Sermage}},
\bibinfo{author}{\bibfnamefont{F.}~\bibnamefont{Bernardot}},
\bibinfo{author}{\bibfnamefont{C.}~\bibnamefont{Testelin}}, and
\bibinfo{author}{\bibfnamefont{M.}~\bibnamefont{Chamarro}},
\bibinfo{journal}{Phys. Rev. B} \textbf{\bibinfo{volume}{75}},
\bibinfo{pages}{205304} (\bibinfo{year}{2007}).

\bibitem[Eble et~al.(2008)Eble, Testelin, Bernardot, Chamarro, and Karczewski]{eble}%
\bibinfo{author}{\bibfnamefont{B.}~\bibnamefont{Eble}},
\bibinfo{author}{\bibfnamefont{C.}~\bibnamefont{Testelin}},
\bibinfo{author}{\bibfnamefont{F.}~\bibnamefont{Bernardot}},
\bibinfo{author}{\bibfnamefont{M.}~\bibnamefont{Chamarro}}, and
\bibinfo{author}{\bibfnamefont{G.}~\bibnamefont{Karczewski}},
\bibinfo{journal}{arXiv:0801.1457v1}  (\bibinfo{year}{2008}).

\bibitem[Chamarro(1998)Chamarro, Bernardot, and Testelin]{chamarro}%
\bibinfo{author}{\bibfnamefont{M.}~\bibnamefont{Chamarro}},
\bibinfo{author}{\bibfnamefont{F.}~\bibnamefont{Bernardot}}, and
\bibinfo{author}{\bibfnamefont{C.}~\bibnamefont{Testelin}},
\bibinfo{journal}{J. Phys.: Consens. Matter} \textbf{\bibinfo{volume}{19}},
\bibinfo{pages}{445007} (\bibinfo{year}{2007}).

\bibitem[Tribollet et~al.(2003)Tribollet, Bernardot, Menant, Karczewski, Testelin,
 and Chamarro]{tribollet1}%
\bibinfo{author}{\bibfnamefont{J.}~\bibnamefont{Tribollet}},
\bibinfo{author}{\bibfnamefont{F.} ~\bibnamefont{Bernardot}},
\bibinfo{author}{\bibfnamefont{M.}~\bibnamefont{Menant}},
\bibinfo{author}{\bibfnamefont{G.}~\bibnamefont{Karczewski}},
\bibinfo{author}{\bibfnamefont{C.}~\bibnamefont{Testelin}}, and
\bibinfo{author}{\bibfnamefont{M.} ~\bibnamefont{Chamarro}},
\bibinfo{journal}{Phys. Rev. B} \textbf{\bibinfo{volume}{68}}, \bibinfo{pages}{235316}  (\bibinfo{year}{2003}).

\bibitem[Astakhov et~al.(2008)Astakhov, Glazov, Yakovlev, Zhukov, Ossau, Molenkamp, and Bayer]%
{astakhov}\bibinfo{author}
\bibinfo{author}{\bibfnamefont{G.~V.}~\bibnamefont{Astakhov}},
\bibinfo{author}{\bibfnamefont{M.~M.}~\bibnamefont{Glazov}},
\bibinfo{author}{\bibfnamefont{D.~R.}~\bibnamefont{Yakovlev}},
\bibinfo{author}{\bibfnamefont{E.~A.}~\bibnamefont{Zhukov}},
\bibinfo{author}{\bibfnamefont{W.}~\bibnamefont{Ossau}},
\bibinfo{author}{\bibfnamefont{L.~W.}~\bibnamefont{Molenkamp}}, and
\bibinfo{author}{\bibfnamefont{M.}~\bibnamefont{Bayer}},
\bibinfo{journal}{Semicond. Sci. Technol.} \textbf{\bibinfo{volume}{23}},
\bibinfo{pages}{114001} (\bibinfo{year}{2008}).

\bibitem[Zhukov(2007)Zhukov, Yakovlev, Bayer, Glazov, Ivchenko, Karczewski, Wojtowicz, and Kossut]{zhukov}%
\bibinfo{author}{\bibfnamefont{E.~A.}~\bibnamefont{Zhukov}},
\bibinfo{author}{\bibfnamefont{D.~R.}~\bibnamefont{Yakovlev}},
\bibinfo{author}{\bibfnamefont{M.}~\bibnamefont{Bayer}},
\bibinfo{author}{\bibfnamefont{M.~M.}~\bibnamefont{Glazov}},
\bibinfo{author}{\bibfnamefont{E.~L.}~\bibnamefont{Ivchenko}},
\bibinfo{author}{\bibfnamefont{G.}~\bibnamefont{Karczewski}},
\bibinfo{author}{\bibfnamefont{T.}~\bibnamefont{Wojtowicz}}, and
\bibinfo{author}{\bibfnamefont{J.}~\bibnamefont{Kossut}},
\bibinfo{journal}{Phys. Rev. B} \textbf{\bibinfo{volume}{76}},
\bibinfo{pages}{205310} (\bibinfo{year}{2007}).

\bibitem[Weng and Wu(2003)Weng and Wu]{weng}%
\bibinfo{author}{\bibfnamefont{M.Q.}~\bibnamefont{Weng}} and
\bibinfo{author}{\bibfnamefont{M.W.}~\bibnamefont{Wu}},
\bibinfo{journal}{Phys.
Rev. B} \textbf{\bibinfo{volume}{68}}, \bibinfo{pages}{075312}  (\bibinfo{year}{2003}).

\bibitem[Zhou et~al(2007)Zhou, Cheng, and Wu]{zhou}%
\bibinfo{author}{\bibfnamefont{J.}~\bibnamefont{Zhou}},
\bibinfo{author}{\bibfnamefont{J.L.}~\bibnamefont{Cheng}} and
\bibinfo{author}{\bibfnamefont{M.W.}~\bibnamefont{Wu}},
\bibinfo{journal}{Phys. Rev. B} \textbf{\bibinfo{volume}{75}},
\bibinfo{pages}{045305}  (\bibinfo{year}{2007}).

\bibitem[Kainz et~al(2004)Kainz, Rossler, and Winkler]{kainz}%
\bibinfo{author}{\bibfnamefont{J.}~\bibnamefont{Kainz}},
\bibinfo{author}{\bibfnamefont{U.}~\bibnamefont{R\"ossler}} and
\bibinfo{author}{\bibfnamefont{R.}~\bibnamefont{Winkler}},
\bibinfo{journal}{Phys. Rev. B} \textbf{\bibinfo{volume}{70}},
\bibinfo{pages}{195322}  (\bibinfo{year}{2004}).

\bibitem[Kavokin(2008)Kavokin]{kavokin2}%
\bibinfo{author}{\bibfnamefont{K.~V.}~\bibnamefont{Kavokin}},
\bibinfo{journal}{Semicond. Sci. Technol.} \textbf{\bibinfo{volume}{23}},
\bibinfo{pages}{114009} (\bibinfo{year}{2008}).

\bibitem[Kavokin(2004)]{kavokin1}%
\bibinfo{author}{\bibfnamefont{K.~V.}~\bibnamefont{Kavokin}},
\bibinfo{journal}{Phys. Rev. B} \textbf{\bibinfo{volume}{69}},
\bibinfo{pages}{075302} (\bibinfo{year}{2004}).

\bibitem[Putikka and Joynt(2004)]{putikka}%
\bibinfo{author}{\bibfnamefont{W.~O.}~\bibnamefont{Putikka}} and
\bibinfo{author}{\bibfnamefont{R.}~\bibnamefont{Joynt}},
\bibinfo{journal}{Phys. Rev. B} \textbf{\bibinfo{volume}{70}},
\bibinfo{pages}{113201} (\bibinfo{year}{2004}).

\bibitem[Harmon(2009)Harmon, Putikka, and Joynt]{harmon}%
\bibinfo{author}{\bibfnamefont{N.~J.}~\bibnamefont{Harmon}},
\bibinfo{author}{\bibfnamefont{W.~O.}~\bibnamefont{Putikka}}, and
\bibinfo{author}{\bibfnamefont{R.}~\bibnamefont{Joynt}},
\bibinfo{journal}{Phys. Rev. B} \textbf{\bibinfo{volume}{79}},
\bibinfo{pages}{115204} (\bibinfo{year}{2009}).

\bibitem[Ghosh et~al.(2005)Ghosh, Sih, Lau, and Awschalom]{ghosh}%
\bibinfo{author}{\bibfnamefont{S.}~\bibnamefont{Ghosh}},
\bibinfo{author}{\bibfnamefont{V.}~\bibnamefont{Sih}},
\bibinfo{author}{\bibfnamefont{W.H.}~\bibnamefont{Lau}}, and
\bibinfo{author}{\bibfnamefont{D.D.} ~\bibnamefont{Awschalom}},
\bibinfo{journal}{Appl. Phys. Lett.} \textbf{\bibinfo{volume}{86}}, \bibinfo{pages}{232507}  (\bibinfo{year}{2005}).

\bibitem[Kikkawa and Awschalom(1998)Kikkawa and Awschalom]{kikkawa}%
\bibinfo{author}{\bibfnamefont{J.M.}~\bibnamefont{Kikkawa}} and
\bibinfo{author}{\bibfnamefont{D.D.}~\bibnamefont{Awschalom}},
\bibinfo{journal}{Phys.
Rev. Lett.} \textbf{\bibinfo{volume}{80}}, \bibinfo{pages}{4313}  (\bibinfo{year}{1998}).

\bibitem[Mahan and Woodworth(2008)Mahan and Woodworth]{mahan}%
\bibinfo{author}{\bibfnamefont{G.D.}~\bibnamefont{Mahan}} and
\bibinfo{author}{\bibfnamefont{R.}~\bibnamefont{Woodworth}},
\bibinfo{journal}{Phys. Rev. B} \textbf{\bibinfo{volume}{78}}, \bibinfo{pages}{075205}  (\bibinfo{year}{2008}).


\bibitem[Kheng et~al.(1993)Kheng, Cox, Merle d'Aubigne, Bassani, Saminadayar,
 and Tatarenko]{kheng}%
\bibinfo{author}{\bibfnamefont{K.}~\bibnamefont{Kheng}},
\bibinfo{author}{\bibfnamefont{R.~T.} ~\bibnamefont{Cox}},
\bibinfo{author}{\bibfnamefont{Y.}~\bibnamefont{Merle d'Aubigne}},
\bibinfo{author}{\bibfnamefont{F.}~\bibnamefont{Bassani}},
\bibinfo{author}{\bibfnamefont{K.}~\bibnamefont{Saminadayar}}, and
\bibinfo{author}{\bibfnamefont{S.} ~\bibnamefont{Tatarenko}},
\bibinfo{journal}{Phys. Rev. Lett.} \textbf{\bibinfo{volume}{71}}, \bibinfo{pages}{1752}  (\bibinfo{year}{1993}).

\bibitem[Adachi et~al.(1997)Adachi, Miyashita, Takeyama, Takagi, and Tackeuchi]{adachi2}%
\bibinfo{author}{\bibfnamefont{S.}~\bibnamefont{Adachi}},
\bibinfo{author}{\bibfnamefont{T.}~\bibnamefont{Miyashita}},
\bibinfo{author}{\bibfnamefont{S.}~\bibnamefont{Takeyama}},
\bibinfo{author}{\bibfnamefont{Y.}~\bibnamefont{Takagi}}, and
\bibinfo{author}{\bibfnamefont{A.} ~\bibnamefont{Tackeuchi}},
\bibinfo{journal}{Journal of Luminescence} \textbf{\bibinfo{volume}{72-74}}, \bibinfo{pages}{307}  (\bibinfo{year}{1997}).

\bibitem[Adachi et~al.(2001)Adachi, Ohno, Matsukura, and Ohno]{adachi}%
\bibinfo{author}{\bibfnamefont{T.}~\bibnamefont{Adachi}},
\bibinfo{author}{\bibfnamefont{Y.}~\bibnamefont{Ohno}},
\bibinfo{author}{\bibfnamefont{F.}~\bibnamefont{Matsukura}}, and
\bibinfo{author}{\bibfnamefont{H.} ~\bibnamefont{Ohno}},
\bibinfo{journal}{Physica E} \textbf{\bibinfo{volume}{10}}, \bibinfo{pages}{36}  (\bibinfo{year}{2001}).


\bibitem[Look(1989)]{look}%
\bibinfo{author}{\bibfnamefont{D.C.}~\bibnamefont{Look}},
\emph{\bibinfo{booktitle}{Electrical Characterization of GaAs Materials and Devices}}%
, (\bibinfo{publisher}{John Wiley \& Sons},
\bibinfo{year}{1989}),  \bibinfo{pages}{p. 113}.


\bibitem[Aschroft(1976)]{ashcroft}%
\bibinfo{author}{\bibfnamefont{N.W.}~\bibnamefont{Ashcroft}} and
\bibinfo{author}{\bibfnamefont{D.}~\bibnamefont{Mermin}},
\emph{\bibinfo{booktitle}{Solid State Physics}}%
,(\bibinfo{publisher}{Thomson Brooks/Cole},
\bibinfo{year}{1976}),  \bibinfo{pages}{p. 581}.

\bibitem[Gor'kov and Krotkov(2003)]{gorkov2}%
\bibinfo{author}{\bibfnamefont{L.~P.}~\bibnamefont{Gor'kov}} and
\bibinfo{author}{\bibfnamefont{P.~L.}~\bibnamefont{Krotkov}},
\bibinfo{journal}{Phys. Rev. B} \textbf{\bibinfo{volume}{67}},
\bibinfo{pages}{033203} (\bibinfo{year}{2003})

\bibitem[Dzyaloshinskii(1958)]{dzyaloshinskii}%
\bibinfo{author}{\bibfnamefont{I.}~\bibnamefont{Dzyaloshinskii}},
\bibinfo{journal}{Phys. Chem. Solids} \textbf{\bibinfo{volume}{4}},
\bibinfo{pages}{241}  (\bibinfo{year}{1958}).

\bibitem[Moriya(1960)]{moriya}%
\bibinfo{author}{\bibfnamefont{T.}~\bibnamefont{Moriya}},
\bibinfo{journal}{Phys. Rev.} \textbf{\bibinfo{volume}{100}},
\bibinfo{pages}{91}  (\bibinfo{year}{1960}).

\bibitem[Ponomarev et~al.(1999)Ponomarev, Flambaum, and Efros]{ponomarev}%
\bibinfo{author}{\bibfnamefont{I.~V.}~\bibnamefont{Ponomarev}},
\bibinfo{author}{\bibfnamefont{V.~V.}~\bibnamefont{Flambaum}}, and
\bibinfo{author}{\bibfnamefont{A.~L.}~\bibnamefont{Efros}},
\bibinfo{journal}{Phys. Rev. B} \textbf{\bibinfo{volume}{60}},
\bibinfo{pages}{5485} (\bibinfo{year}{1999}).

\bibitem[Merkulov et~al.(2002)Merkulov, Efros, and Rosen]{merkulov}%
\bibinfo{author}{\bibfnamefont{I.~A.}~\bibnamefont{Merkulov}},
\bibinfo{author}{\bibfnamefont{A.~L.}~\bibnamefont{Efros}}, and
\bibinfo{author}{\bibfnamefont{M.}~\bibnamefont{Rosen}},
\bibinfo{journal}{Phys. Rev. B} \textbf{\bibinfo{volume}{65}},
\bibinfo{pages}{205309} (\bibinfo{year}{2002}).

\bibitem[Braun et~al.(2005)Braun, Marie, Lombez, Urbaszek, Amand, Renucci,Kalevich, Kavokin, Krebs, Voisin
 and Masumoto]{braun}%
\bibinfo{author}{\bibfnamefont{P.~F.}~\bibnamefont{Braun}},
\bibinfo{author}{\bibfnamefont{X.} ~\bibnamefont{Marie}},
\bibinfo{author}{\bibfnamefont{L.}~\bibnamefont{Lombez}},
\bibinfo{author}{\bibfnamefont{B.}~\bibnamefont{Urbaszek}},
\bibinfo{author}{\bibfnamefont{T.} ~\bibnamefont{Amand}},
\bibinfo{author}{\bibfnamefont{P.}~\bibnamefont{Renucci}},
\bibinfo{author}{\bibfnamefont{V.~K.}~\bibnamefont{Kalevich}},
\bibinfo{author}{\bibfnamefont{K.~V.}~\bibnamefont{Kavokin}},
\bibinfo{author}{\bibfnamefont{O.}~\bibnamefont{Krebs}},
\bibinfo{author}{\bibfnamefont{P.}~\bibnamefont{Voisin}}, and
\bibinfo{author}{\bibfnamefont{Y.} ~\bibnamefont{Masumoto}},
\bibinfo{journal}{Phys. Rev. Lett.} \textbf{\bibinfo{volume}{94}}, \bibinfo{pages}{116601}  (\bibinfo{year}{2005}).

\bibitem[Dyakonov(1974)]{dyakonov1}%
\bibinfo{author}{\bibfnamefont{M.~I.}~\bibnamefont{Dyakonov}} and
\bibinfo{author}{\bibfnamefont{V.~I.}~\bibnamefont{Perel}},
\bibinfo{journal}{Sov. Phys. JETP} \textbf{\bibinfo{volume}{33}},
\bibinfo{pages}{1053} (\bibinfo{year}{1974}).

\bibitem[Dyakonov(1986)]{dyakonov2}%
\bibinfo{author}{\bibfnamefont{M.~I.}~\bibnamefont{Dyakonov}} and
\bibinfo{author}{\bibfnamefont{V.~Y.}~\bibnamefont{Kachorovskii}},
\bibinfo{journal}{Sov. Phys. Semicond.} \textbf{\bibinfo{volume}{20}},
\bibinfo{pages}{178} (\bibinfo{year}{1986}).

\bibitem[Fu and We(2008)Fu and Wu]{fu}%
\bibinfo{author}{\bibfnamefont{J.Y.}~\bibnamefont{Fu}} and
\bibinfo{author}{\bibfnamefont{M.W.}~\bibnamefont{Wu}},
\bibinfo{journal}{J. Appl. Phys.} \textbf{\bibinfo{volume}{104}}, \bibinfo{pages}{093712}  (\bibinfo{year}{2008}).

\bibitem[Rashba(1960)]{rashba}%
\bibinfo{author}{\bibfnamefont{E.~I.}~\bibnamefont{Rashba}},
\bibinfo{journal}{Sov. Phys. Solid State} \textbf{\bibinfo{volume}{2}},
\bibinfo{pages}{1109} (\bibinfo{year}{1960}).

\bibitem[Ohno et~al.(2000)Ohno, Terauchi, Adachi, Matsukura, and Ohno]{ohno}%
\bibinfo{author}{\bibfnamefont{Y.}~\bibnamefont{Ohno}},
\bibinfo{author}{\bibfnamefont{R.}~\bibnamefont{Terauchi}},
\bibinfo{author}{\bibfnamefont{T.}~\bibnamefont{Adachi}},
\bibinfo{author}{\bibfnamefont{F.}~\bibnamefont{Matsukura}}, and
\bibinfo{author}{\bibfnamefont{H.}~\bibnamefont{Ohno}},
\bibinfo{journal}{Physica E} \textbf{\bibinfo{volume}{6}}, \bibinfo{pages}{817}  (\bibinfo{year}{2000}).


\bibitem[Hassenkam et~al.(1997)Hassenkam, Pedersen, Baklanov, Kristensen, Sorensen, Lindelof, Pikus, and Pikus]{hassenkam}%
\bibinfo{author}{\bibfnamefont{T.}~\bibnamefont{Hassenkam}},
\bibinfo{author}{\bibfnamefont{S.}~\bibnamefont{Pedersen}},
\bibinfo{author}{\bibfnamefont{K.}~\bibnamefont{Baklanov}},
\bibinfo{author}{\bibfnamefont{A.}~\bibnamefont{Kristensen}},
\bibinfo{author}{\bibfnamefont{C.B.}~\bibnamefont{Sorensen}}, 
\bibinfo{author}{\bibfnamefont{P.E.} ~\bibnamefont{Lindelof}},
\bibinfo{author}{\bibfnamefont{F.G.}~\bibnamefont{Pikus}},and
\bibinfo{author}{\bibfnamefont{G.E.}~\bibnamefont{Pikus}},
\bibinfo{journal}{Phys. Rev. B} \textbf{\bibinfo{volume}{55}}, \bibinfo{pages}{9298}  (\bibinfo{year}{1997}).

\bibitem[Elliott(1954)]{elliott}%
\bibinfo{author}{\bibfnamefont{R.~J.}~\bibnamefont{Elliott}},
\bibinfo{journal}{Phys. Rev.} \textbf{\bibinfo{volume}{96}},
\bibinfo{pages}{266} (\bibinfo{year}{1954}).

\bibitem[Yafet(1963)]{yafet}%
\bibinfo{author}{\bibfnamefont{Y.}~\bibnamefont{Yafet}}, in
\emph{\bibinfo{booktitle}{Solid State Physics}}, edited by
\bibinfo{editor}{\bibfnamefont{F.}~\bibnamefont{Seitz}} and
\bibinfo{editor}{\bibfnamefont{D.}~\bibnamefont{Turnbull}}
(\bibinfo{publisher}{Academic, New York}, \bibinfo{year}{1963}),
\bibinfo{volume}{Vol. 14},
\bibinfo{pages}{pp. 1-63}.

\bibitem[Chazalviel(1975)Chazalviel]{chazalviel}%
\bibinfo{author}{\bibfnamefont{J.~N.}~\bibnamefont{Chazalviel}},
\bibinfo{journal}{Phys.
Rev. B} \textbf{\bibinfo{volume}{11}}, \bibinfo{pages}{1555}  (\bibinfo{year}{1975}).

\bibitem[Ridley(2009)]{ridley}%
\bibinfo{author}{\bibfnamefont{B.~K.}~\bibnamefont{Ridley}},
\emph{\bibinfo{booktitle}{Electrons and Phonons in Semiconductor Multilayers}}%
,\bibinfo{edition}{~2nd ed.~}  (\bibinfo{publisher}{Cambridge University Press, New York},
\bibinfo{year}{2009}),  \bibinfo{pages}{pp. 313, ~342}.

\bibitem[Tackeuchi et~al.(1999)Tackeuchi, Kuroda, Muto, Nishikawa, and Wada]{tackeuchi}%
\bibinfo{author}{\bibfnamefont{A.}~\bibnamefont{Tackeuchi}},
\bibinfo{author}{\bibfnamefont{T.}~\bibnamefont{Kuroda}},
\bibinfo{author}{\bibfnamefont{S.}~\bibnamefont{Muto}},
\bibinfo{author}{\bibfnamefont{Y.}~\bibnamefont{Nishikawa}}, and
\bibinfo{author}{\bibfnamefont{O.} ~\bibnamefont{Wada}},
\bibinfo{journal}{Jpn. J. Appl. Phys.} \textbf{\bibinfo{volume}{38}}, \bibinfo{pages}{4680}  (\bibinfo{year}{1999}).

\bibitem[Bir et~al.(1976)Bir, Aronov, and Pikus]{bir}%
\bibinfo{author}{\bibfnamefont{G.~L.}~\bibnamefont{Bir}},
\bibinfo{author}{\bibfnamefont{A.~G.}~\bibnamefont{Aronov}}, and
\bibinfo{author}{\bibfnamefont{G.~E.}~\bibnamefont{Pikus}},
\bibinfo{journal}{Sov. Phys. JETP} \textbf{\bibinfo{volume}{42}},
\bibinfo{pages}{705} (\bibinfo{year}{1976}).

\bibitem[Song and Kim(2002)Song and Kim]{song}%
\bibinfo{author}{\bibfnamefont{Pil Hun}~\bibnamefont{Song}} and
\bibinfo{author}{\bibfnamefont{K.W.}~\bibnamefont{Kim}},
\bibinfo{journal}{Phys.
Rev. B} \textbf{\bibinfo{volume}{66}}, \bibinfo{pages}{035207}  (\bibinfo{year}{2002}).

\bibitem[Ohno et~al.(1999)Ohno, Terauchi, Adachi, Matsukura, and Ohno]{ohno2}%
\bibinfo{author}{\bibfnamefont{Y.}~\bibnamefont{Ohno}},
\bibinfo{author}{\bibfnamefont{R.}~\bibnamefont{Terauchi}},
\bibinfo{author}{\bibfnamefont{T.}~\bibnamefont{Adachi}},
\bibinfo{author}{\bibfnamefont{F.}~\bibnamefont{Matsukura}}, and
\bibinfo{author}{\bibfnamefont{H.} ~\bibnamefont{Ohno}},
\bibinfo{journal}{Phys. Rev. Lett.} \textbf{\bibinfo{volume}{83}}, \bibinfo{pages}{4196}  (\bibinfo{year}{1999}).

\bibitem[Harrison(2005)]{footnote}%
\bibinfo{}{
We note that the experimental rates of 
Ohno et al. depicted here are twice that of what is reported in the actual experiments by that group. This is due to different definitions of spin relaxation time.\cite{kainz}}

\bibitem[Harrison(2005)]{harrison}%
\bibinfo{author}{\bibfnamefont{P.}~\bibnamefont{Harrison}},
\emph{\bibinfo{booktitle}{Quantum Wells, Wires and Dots}}%
,\bibinfo{edition}{~2nd ed.~}  (\bibinfo{publisher}{Wiley \& Sons, West Sussex},
\bibinfo{year}{2005}),  \bibinfo{pages}{p. 137}.

\bibitem[Davies(1998)]{davies}%
\bibinfo{author}{\bibfnamefont{J.}~\bibnamefont{Davies}},
\emph{\bibinfo{booktitle}{The Physics of Low Dimensional Semiconductors}}%
, (\bibinfo{publisher}{Cambridge University Press, New York},
\bibinfo{year}{1998}),  \bibinfo{pages}{p. 86}.

\bibitem[Schliemann et~al(2003)Schliemann, Khaetskii, and Loss]{schliemann}%
\bibinfo{author}{\bibfnamefont{J.}~\bibnamefont{Schliemann}},
\bibinfo{author}{\bibfnamefont{A.}~\bibnamefont{Khaetskii}} and
\bibinfo{author}{\bibfnamefont{D.}~\bibnamefont{Loss}},
\bibinfo{journal}{J. Phys.: Condens. Matter} \textbf{\bibinfo{volume}{15}},
\bibinfo{pages}{1809}  (\bibinfo{year}{2003}).

\bibitem[Dohrmann et~al.(2004)Dohrmann, Hagele, Rudulph, Bichler, Schuh, and Oestreich]{dohrmann}%
\bibinfo{author}{\bibfnamefont{S.}~\bibnamefont{Dohrmann}},
\bibinfo{author}{\bibfnamefont{D.}~\bibnamefont{Hagele}},
\bibinfo{author}{\bibfnamefont{J.}~\bibnamefont{Rudulph}},
\bibinfo{author}{\bibfnamefont{M.}~\bibnamefont{Bichler}},
\bibinfo{author}{\bibfnamefont{D.}~\bibnamefont{Schuh}}, and
\bibinfo{author}{\bibfnamefont{M.} ~\bibnamefont{Oestreich}},
\bibinfo{journal}{Phys. Rev. Lett.} \textbf{\bibinfo{volume}{93}}, \bibinfo{pages}{147405}  (\bibinfo{year}{2004}).

\bibitem[Zaitsev et~al.(2007)Zaitsev, Welsch, Forchel, and Bacher]{zaitsev}%
\bibinfo{author}{\bibfnamefont{S.~V.}~\bibnamefont{Zaitsev}},
\bibinfo{author}{\bibfnamefont{M.~K.}~\bibnamefont{Welsch}},
\bibinfo{author}{\bibfnamefont{A.}~\bibnamefont{Forchel}}, and
\bibinfo{author}{\bibfnamefont{G.} ~\bibnamefont{Bacher}},
\bibinfo{journal}{Journal of Experimental and Theoretical Physics} \textbf{\bibinfo{volume}{105}}, \bibinfo{pages}{1241}  (\bibinfo{year}{2007}).

\bibitem[Segall et~al.(1963)Segall, Lorenz, and Halsted]{segall}%
\bibinfo{author}{\bibfnamefont{B.}~\bibnamefont{Segall}},
\bibinfo{author}{\bibfnamefont{M.R.}~\bibnamefont{Lorenz}}, and
\bibinfo{author}{\bibfnamefont{R.E.} ~\bibnamefont{Halsted}},
\bibinfo{journal}{Phys. Rev.} \textbf{\bibinfo{volume}{129}}, \bibinfo{pages}{2471}  (\bibinfo{year}{1963}).


\end{thebibliography}

\end{document}